\documentclass[preprint,prb,twoside,amsmath,aps]{revtex4}

\usepackage{graphicx}

\begin{document}

\title{Spatial dispersion of the high-frequency conductivity of two-dimensional electron gas subjected to a high electric field: collisionless case}

\author{V. V. Korotyeyev \footnote{koroteev@ukr.net. This article may be downloaded for personal use only. Any other use requires prior permission of the author and AIP Publishing. This article appeared in (Appl. Phys. Lett. 113, 041102 (2018)) and may be found at (https://doi.org/10.1063/1.5041322)}}
\affiliation{Department of Theoretical Physics, Institute of Semiconductor Physics of NAS of Ukraine,  03028 Kyiv,
Ukraine}
\author{V. A. Kochelap}
\affiliation{Department of Theoretical Physics, Institute of Semiconductor Physics of NAS of Ukraine,  03028 Kyiv,
Ukraine}
\author{S. Danylyuk}
\affiliation{Chair for the Technology of Optical Systems, RWTH Aachen University,  52074 Aachen, Germany}
\author{L. Varani}
\affiliation{Institute of Electronics and Systems, UMR CNRS 5214, University of Montpellier,  France}

\begin{abstract}
We present the analysis of high-frequency (dynamic) conductivity with the
spatial dispersion, $\sigma (\omega, {\bf q})$, of two-dimensional electron gas subjected to a high electric
 field. We found that at finite wavevector, ${\bf q}$,  and at high fields, the
 high-frequency conductivity shows  following peculiarities: strong non-reciprocal dispersion; oscillatory behavior;
 a  set of frequency regions with negative $\sigma'$; non-exponential
 decay of $\sigma'$ and $\sigma''$ with  frequency (opposite to the
  Landau damping mechanism). We illustrate the general results by calculations of
 spectral characteristics of particular plasmonic heterostructures on the basis
 of III-V semiconductor compounds. We
  conclude that the detailed analysis of the spatial dispersion of the dynamic
  conductivity of 2DEG subjected to high electric fields is critically important
  for different THz applications.

\end{abstract}

\maketitle

The high-frequency properties of a two-dimensional electron gas (2DEG) are determined
by the dynamic conductivity, $\sigma (\omega, {\bf q})$,
where $\omega$ and $\bf q$ are the angular frequency and the wavevector of the electric field,
$E_{\omega, \bf{ q}}$, of the electromagnetic wave.
The dependence of $\sigma (\omega, {\bf q})$ on the wavevector,
i.e., the spatial dispersion, becomes especially important
for samples with submicron- and nanosized lateral structuring.
Indeed, a plane electromagnetic wave illuminating a laterally nonuniform
sample induces electric field components varying both in space and time,
which interact with the 2DEG. The spatial dependence of these field components  is defined by
characteristic scales of the lateral structuring of the sample.

Examples of laterally nonuniform structures with 2DEGs
are grating-gated structures, surface-relief grating,
plasmonic and metamaterial nanodevices based
on the excitation of 2D plasmon modes, etc.\cite{Ho}
These structures can be used for different applications, including
detecting and emitting devices of far-infrared and terahertz radiation. In particular,  amplification
of charge oscillations with sub-micron wavelengths
is possible in these structures at high electric fields\cite{THz-Ampl-1,THz-Ampl-2,THz-Ampl-22,THz-Ampl-3,THz-Ampl-4}
(further references, including early papers can be found in Ref.~[\onlinecite{THz-Ampl-5}]).

Recently \cite{Nano-opt-1, Nano-opt-2, Nano-opt-3}, novel near-field optical microscope techniques
have been proposed, where electric fields  (varying in time and space)  are excited
at the length scale of tens of nanometers. The development of such methods
facilitates the exploration of excitations and fields  at the short time and  length scales.
In all presented examples, a detailed analysis of
the spatial dispersion of the dynamic conductivity is critically
important.

In the case of samples with submicron-- or nano--scaled lateral structurization,
when  the characteristic lateral scales are shorter than
the mean free path of the electrons, $l_{sc}$, and frequencies are greater than the inverse scattering time, $1/\tau_{sc}$, the dynamic conductivity should be found by solving the Boltzmann transport equation (BTE) in collisionless (ballistic) approximation.
For this case, $\sigma (\omega, {\bf q})$ has
both real, $\sigma'$, and imaginary, $\sigma''$, parts.
The non-zero real part $\sigma'$ is due to the strong phase-mixing property of the BTE~\cite{Math-1},
which leads also to the well known Landau damping mechanism for charge oscillations of
an equilibrium electron gas\cite{Landau,Lifshitz}.
When the electrons are drifting in an electric
field, the effect of this field on the conductivity is typically taken into account
only by using the so-called shifted Maxwellian distribution which ignores the
effect of the electric field on electron high-frequency dynamics.
This approach corresponds to
the case, when the term proportional to the stationary  electric field,  $E_0$,
is omitted in the BTE formulated for the high-frequency contribution to the distribution
function. However, a  number of THz applications requires the use of high (lateral)
electric fields applied to the above discussed laterally
nonuniform structures.

In this letter, we present an analysis of the high-frequency conductivity with the
spatial dispersion, $\sigma (\omega, {\bf q})$, for 2DEG subjected to a high stationary electric
 field,  keeping the non-zero electric field term in the BTE. We solved
 the BTE in the collisionless limit and calculated the dynamic  conductivity.
 We found that the effect of the field $E_0$ is significant if
 the relative gain of electron energy from the field $E_0$ for a spatial period
 of the electromagnetic wave is of the order of $1$. For the case, when hot electrons
 can be characterized by the electron temperature, $T_e$, this parameter
 is  $\gamma_{q} = e E_0/qk_{B}T_{e}$, where,  $e$ is the elementary charge and
 $k_B$  is the Boltzmann constant.
 At a given $q$, and high field, $E_0$, the dynamic conductivity
 shows the following peculiarities: (i) oscillatory behavior versus frequency; (ii) a
 set of frequency regions with negative $\sigma''$; (iii) non-exponential
 decay of $\sigma'$ and $\sigma''$ with the frequency (opposite to the
  Landau damping mechanism). We illustrate these general results by calculations of the spectral characteristics of
 particular plasmonic heterostructures based on a III-V semiconductor compounds.

In the frame of the semiclassical analysis, the electron characteristics can
be calculated using the electron distribution function $G({\bf p},{\bf r},t)$ of 2DEG, which, in general, depends on the electron momentum, ${\bf p}$,  the coordinate vector, $ {\bf r}$, and time, $t$. $G({\bf p},{\bf r},t)$ is the solution of the BTE:
\begin{equation}
\frac{\partial G}{\partial t}+{\bf v}\frac{\partial G}{\partial {\bf r}}
-e {\bf E}({\bf r},t)\frac{\partial G}{\partial {\bf p}}=\hat{I}\{ G\},
\label{Boltzmann}
\end{equation}
where ${\bf E} ({\bf r},t) = {\bf E}_0 + \tilde {\bf E} ({\bf r}, t)$
is the total electric field given by the sum of the stationary field ${\bf E}_0$ and $\tilde {\bf E} ({\bf r}, t) = \tilde{\bf E}_{\omega, {\bf q}} e^{i{\bf q}{\bf r}-i(\omega + i \delta) t}$
representing the field associated with the electromagnetic wave and $\hat{I}\{ G\}$ is the collision integral.
In the expression for $\bf \tilde E$, we introduced a parameter $\delta\rightarrow +0$,
which corresponds to adiabatically-slow  turning--on of this field at
$t\rightarrow -\infty$\cite{Lifshitz}.
The total distribution function can be presented as
$G({\bf p},{\bf r},t) = g_0 ({\bf p}) + \tilde{g}({\bf p},{\bf r},t)$,
where  $g_0$ is the stationary distribution function of the electrons in the field ${\bf E}_0$ and
$\tilde{g}({\bf p},{\bf r},t) = \tilde{g}_{\omega,{\bf q}}({\bf p}) e^{i{\bf q}{\bf r}-i(\omega + i \delta)t}$
represents the  time-- and space--dependent perturbation of the distribution function induced by the field $\tilde {\bf E} ({\bf r}, t)$.
We apply our analysis for electrons assuming a parabolic dispersion law and an effective mass, $m^{*}$.

Assuming that the amplitude of the wave field, $\tilde{\bf E}_{\omega, {\bf q}}$, is small,
we can write down the equations for $g_0$ and $\tilde{g}_{\omega,{\bf q}}$:
   \begin{eqnarray}
   \label{ini}
-e {\bf E}_{0}\frac{\partial g_{0}}{\partial {\bf p}}&=&\hat{I}\{ g_{0}\},\\
-i\left(\omega+i\delta-\frac{{\bf q}{\bf p}}{m^{*}}\right)\tilde{g}_{\omega, {\bf q}} - e {\bf E}_{0}
 \frac{\partial \tilde{g}_{\omega, {\bf q}}}{\partial {\bf p}}&=&
 e \tilde{\bf E}_{\omega, {\bf q}}\frac{\partial g_{0}}{\partial{\bf p}},
\label{ini1}
\end{eqnarray}
where Eq.(\ref{ini1}) is written down in collisionless approximation.
Let the electrons be confined in a thin plane layer, say in the $\{x,y\}$-plane.
Then, $\bf p$ and $\bf r$ are two-dimensional vectors.  For the
electric field components, which appeared in Eqs.(\ref{ini}) and (\ref{ini1}), we assume
${\bf E}_{0}=\{-E_{0},0\}$ and ${\bf \tilde E}_{\omega, \bf q}=\{\tilde{E}_{\omega, \bf q},0\}$ with ${\bf q}=\{q,0\}$.

 For high-frequency perturbation of the distribution function, $\tilde{g}_{\omega,q}$,
we solved Eq.~(\ref{ini1}). The solution  satisfying
the condition $\tilde{g}_{\omega,q} \rightarrow 0$ at $p_{x,y}\rightarrow\pm\infty$ is
\begin{eqnarray}
\tilde{g}_{\omega, {q}}=\frac{\tilde{E}_{\omega, {q}}}{E_{0}}
\int_{-\infty}^{p_{x}}dp'_{x}\exp\left[\frac{i}{eE_{0}}\left[(\omega + i \delta)
(p_{x}-p'_{x})\right.\right. \nonumber \\
\left.\left. -\frac{q_{x}(p_{x}^2-{p'}_{x}^{2})}{2m^{*}}\right]\right]\frac{\partial g_{0}(p'_{x},p_{y})}{\partial p'_{x}}\,.
\label{distr_ac}
\end{eqnarray}
Having the function $\tilde{g}_{\omega, {q}}(p_x,p_y)$, we can calculate
the alternative current,
$\tilde{j}_{\omega, {q}}=-e/m^{*}\times\int dp_x dp_y p_x \tilde{g}_{\omega, { q}}$,  and
define the high-frequency conductivity as
$\sigma_{\omega,q}=\tilde{j}_{_{\omega,q}}/\tilde{E}_{\omega, q}$.
We found that $\sigma_{\omega,q}$ can be expressed
in the form of a single integral:
\begin{eqnarray}
\sigma_{\omega,q}=-\frac{e^2n_{0}\omega}{qeE_{0}}\left[1-\sqrt{\frac{i\pi q}{2m^{*}eE_{0}}}\int_{-\infty}^{+\infty}dp_{x}
p_{x}\right. \nonumber\\
\times\left.{\cal W}\left[i\sqrt{\frac{iq}{2m^{*}eE_{0}}}p_{x}\right]\bar{g}_{0}\left[p_{x}+\frac{m^{*}\omega}{q}\right]\right]
\label{sigma_full}
\end{eqnarray}
where ${\cal W}[\xi]$ is the so-called the plasma dispersion
function \cite{Temme}, the function
$\bar{g}_{0}(p_{x}) =\int dp_y g_0(p_x,p_y)/\int dp_xdp_y g_0(p_x,p_y)$
is normalized stationary electron distribution dependent on
the momentum $p_x$ along field direction.
This result can be applied for any form of  the stationary
distribution function, $g_{0}$, with rapid (e.g., exponential) decrease
at large momenta. The latter allows us to set $\delta=0$ in Eq.~(\ref{sigma_full}).
Noteworthy, the function $g_{0}$ is a solution of Eq.~(\ref{ini}), in which
all actual collision processes should be taken into account.
Examples of such functions can be found elsewhere \cite{Ferry,Kor0, Mosko}.

The detail analysis of the $\sigma_{\omega,q}$, we will perform for the so-called shifted Maxwellian function,
\begin{equation}
g_{0}  =\frac{n_{0}}{2\pi m^{*}k_{B}T_{e}}\exp\left[-\frac{(p_{x}-m^{*}V_{dr})^2+p_y^{2}}{2m^{*}k_{B}T_{e}}\right],
\label{g_0}
\end{equation}
 where $n_{0}$, $V_{dr}$ and $T_{e}$ are the electron concentration,
 drift velocity, and electron temperature, respectively.
 For this function, the main peculiarities of $\sigma_{\omega,q}$
 including the affect of the high electric field can be studied analytically.
The parameters $V_{dr}$ and $T_{e}$ are functions
of the applied field, $E_0$, and can be found using the momentum and energy balance
equations (see, for example Refs.~[\onlinecite{Kor1}, \onlinecite{Kor2}]).
Using $g_0(p_x,p_y)$
from Eq.~(\ref{g_0}) and performing integrations, we obtain $\sigma_{\omega,q}$ in the following form:
\begin{eqnarray}
\sigma_{\omega,q} =  -i \frac{2 e^2 \omega n_0}{k_B T_e q^2}
\int_0^{\infty} dx\, x \exp{\left[2 i x \frac{\omega - q V_{dr}}{|q| V_T}\right]} \nonumber\\
\times\exp{\left[- x^2 \left(1 + i\frac{e E_0}{q k_B T_e}\right)\right]}\,,
\label{sigma1}
\end{eqnarray}
with $V_T=\sqrt{2 k_B T_e/m^*} $ being the thermal velocity of the electrons. Finally,
Eq.~(\ref{sigma1}) can be rewritten in terms of the
the plasma dispersion function:
\begin{eqnarray}
\sigma_{\omega, q}= - i \frac{e^2 \omega n_0}{k_B T_e q^2(1+i \gamma_q)}
\left[1 + \frac{\omega - q V_{dr}}{|q| V_T \sqrt{1+i \gamma_q}}\right.\nonumber\\
\times{\cal W} \left.\left[\frac{\omega - q V_{dr}}{|q| V_T \sqrt{1+i \gamma_q} }\right] \right]\,.
 \label{sigma2}
\end{eqnarray}

As expected, $\sigma_{\omega,q}$ depends on the field, $E_0$, via two
parameters of the stationary distribution function of Eq.~(\ref{g_0}),
$T_e, V_{dr}$ and the parameter $\gamma_q$  that is dependent on the wavevector of the electromagnetic field.

If the $\gamma_q$ is negligibly small, we obtain the real part of the dynamic conductivity in the
form\cite{Remark-1}:
\begin{equation} \label{sigma3}
\sigma_{\omega, q}' = \frac{\sqrt{\pi}e^2 n_0 \, \omega}{k_B T_e q^2}\,
\frac{\omega - q V_{dr}}{|q| V_T }\,
\exp{ \left[-\left(\frac{\omega - q V_{dr}}{|q| V_T } \right)^2\right]}.
\end{equation}
Noteworthy, $\sigma_{\omega, q}'$ presented by Eq.~(\ref{sigma3}) is negative for
$\omega < V_{dr}q$, i.e., under this Cerenkov-like condition, the drifted electrons return their energy
to the electromagnetic wave. At $\omega > V_{dr}q$, $\sigma_{\omega, q}'$ is always positive,
it  reaches a maximum and then decreases exponentially
in the high frequency range. For the same limit,  $\gamma_q \rightarrow 0$, the imaginary
part of the conductivity is given by
\begin{equation} \label{sigma4}
\sigma_{\omega,q}'' =- \frac{e^2 n_0 \omega}{k_B T_e q^2}\,\left[1- 2\frac{\omega - q V_{dr}}{|q| V_T} {\cal D}
\left[\frac{\omega - q V_{dr}}{|q| V_T } \right] \right],
\end{equation}
with ${\cal D} [\xi]$ being the Dawson function\cite{Temme}. When the factor
$\xi=(\omega - q V_{dr})/|q| V_T$ is large, we obtain $\sigma_{\omega,q}'' \approx
e^2 n_0/m^*\times\omega/(\omega-q V_{dr})^2$, that corresponds to the response of
the 2DEG in the hydrodynamic limit, i.e., $T_e \rightarrow 0$\cite{THz-Ampl-2, Ecker}.

Returning to the case $\gamma_{q}\neq 0$, the determination of the real and imaginary parts
of $\sigma_{\omega,q}$ given by Eq.~(\ref{sigma2}) requires numerical calculations. However, keeping only terms of the first
order with respect to $\gamma_{q}$  we obtain at small $\gamma_{q}$:
\begin{eqnarray}
  \sigma_{\omega, q}' =\frac{e^2 n_0 \omega}{k_B T_e q^2}
 \left[\sqrt{\pi}\xi\exp(-\xi^2)  -\gamma_{q}\left(1-\xi^2 \right.\right.\nonumber\\
 -\left.\left.(3\xi-2\xi^3){\cal D}(\xi)\right)\right].
 \label{sigma5}
 \end{eqnarray}
From this result we find that at small $\xi$,
$\sigma_{\omega, q}'\propto \sqrt{\pi}\xi-\gamma_{q}$. Thus, the real part of the conductivity
changes the sign at $\xi_{C}=\gamma_{q}/\sqrt{\pi}$, i.e., the Cerenkov-like
instability region is wider than that predicted by Eq.~(\ref{sigma3}) and it is realized when the condition
$0 < \omega < V_{dr} q + {2 e E_0}/{\sqrt{\pi} m^* V_T}$ is fulfilled.
Additionally, from Eq.~(\ref{sigma5}) it follows that at $\xi \gg 1$ the  term
proportional to $\gamma_{q}$ dominates and $\sigma_{\omega, q}'$
is of well-defined negative sign. This result indicates that, at large $\omega$,
there is at least one {\it additional} region of instability.
We notice that when $qV_{dr} <0$ the real part of $\sigma_{\omega, q}$ is always positive.
The imaginary part of the conductivity, $\sigma_{\omega,q}''$,
is weakly modified  at small $\gamma_{q}$.
\begin{figure}[h]
\centering
    \includegraphics[width=0.45\textwidth]{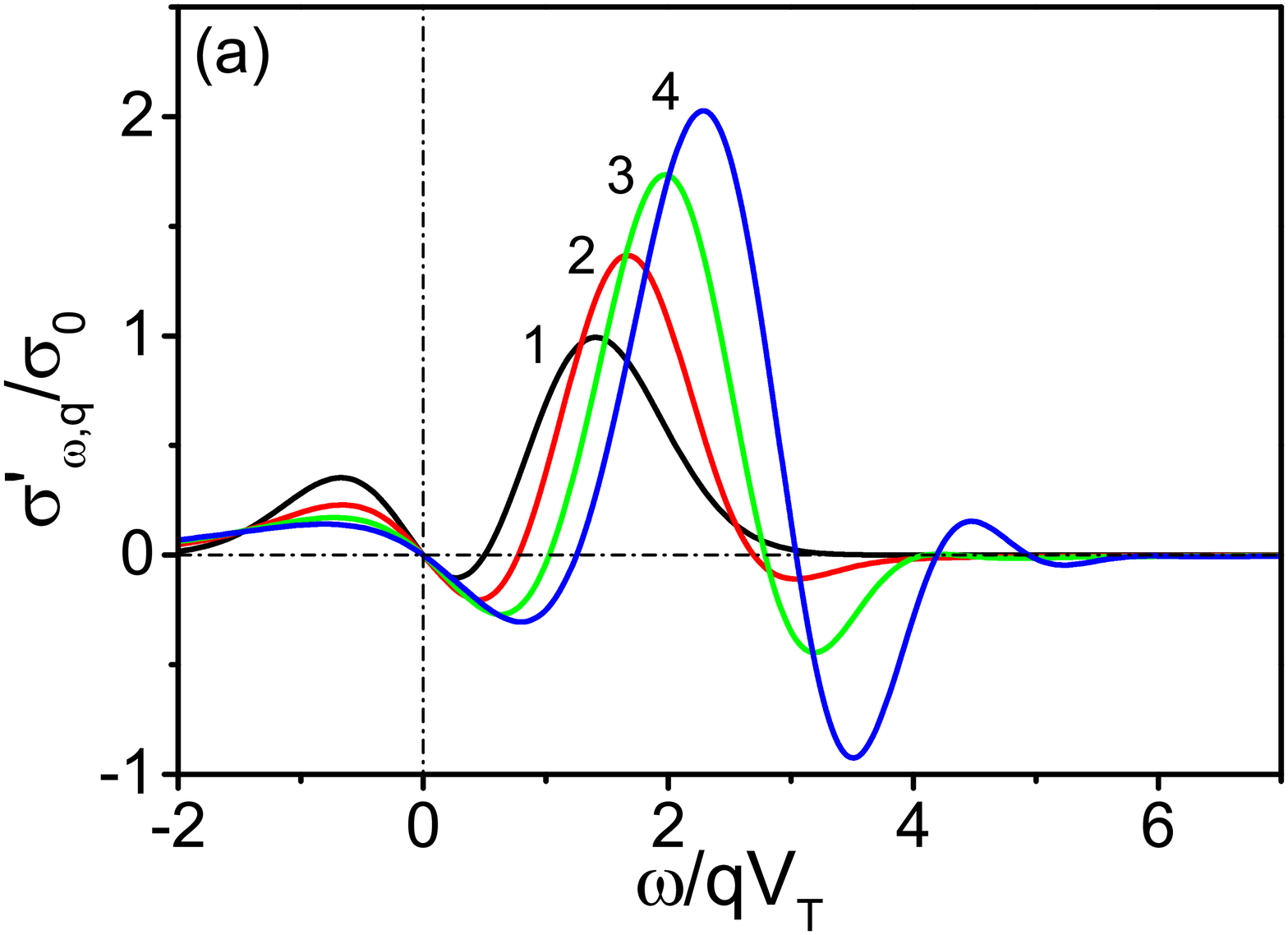}
    \includegraphics[width=0.45\textwidth]{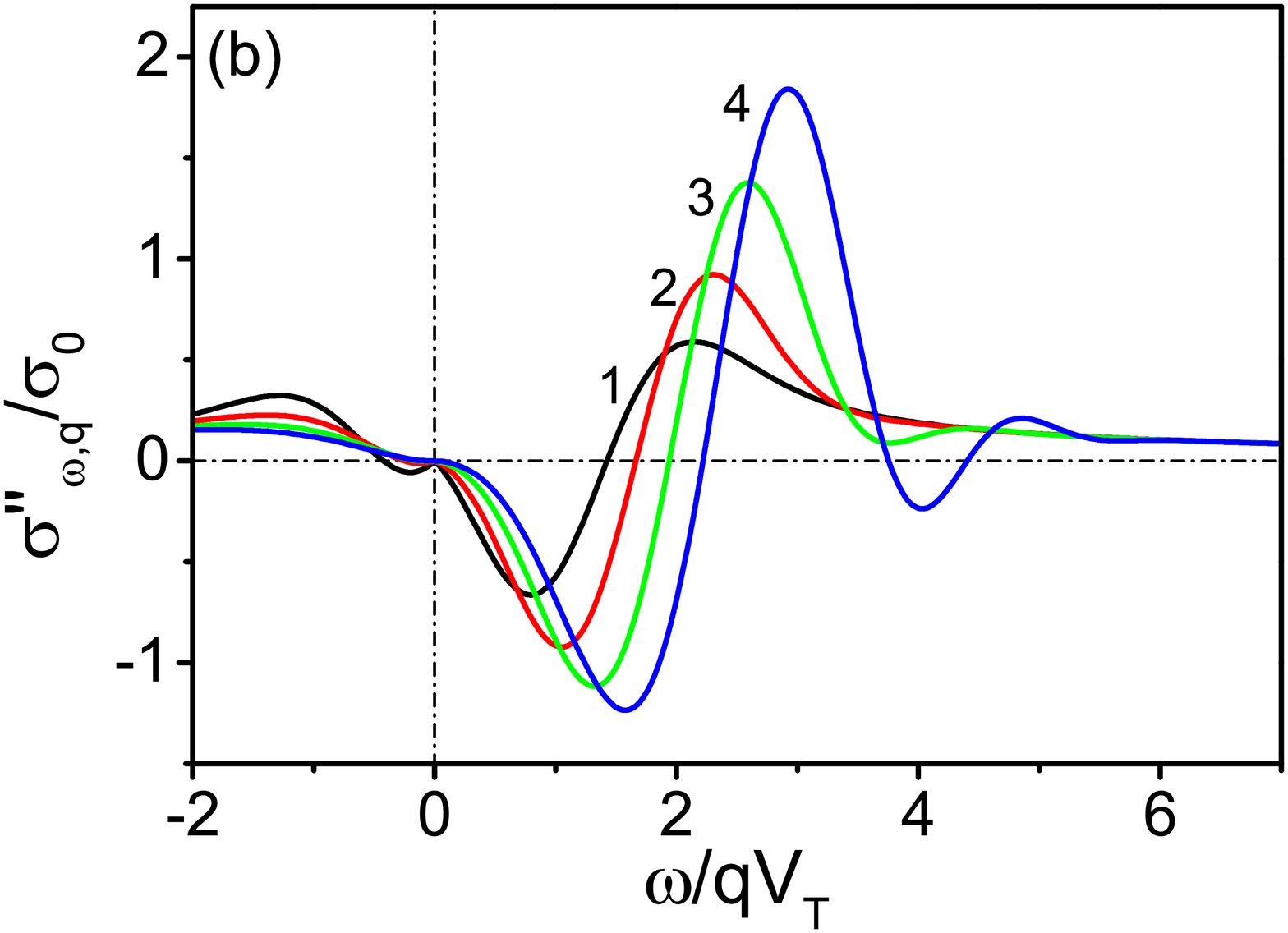}
  \caption{Normalized real (a) and imaginary (b) parts of the dynamic
  conductivity as functions of the normalized angular frequency at $V_{dr}/V_T =0.5$.
  The vertical dot-dashed line divides each figure into two parts: the right part
  is for $q>0$, the left one is for $q < 0$. Curves $1,\,2,\,3$ and $4$ are for $|\gamma_q| = 0,\,0.5,\,1$ and $1.5$, respectively. }
  \label{fig1}
\end{figure}

\begin{figure}[h]
\centering
\includegraphics[width=0.45\textwidth]{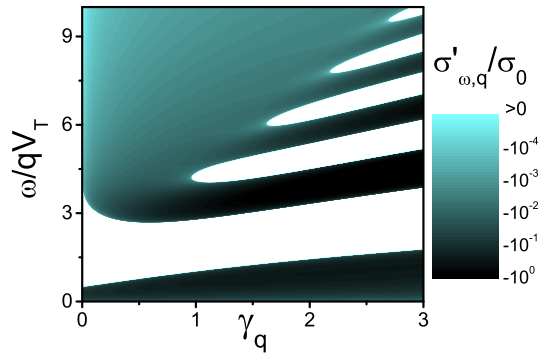}
\caption{Density plot of the normalized $\sigma_{\omega,q}'$ as a function of the dimensionless
frequency $\omega/q V_T$ and dimensionless electric field $\gamma_q$. }
\label{fig2}
\end{figure}

The conclusions  obtained from analytical considerations are supported by
the numerical results shown in Figs.~\ref{fig1} and \ref{fig2}.
Indeed Figs.~\ref{fig1} (a) and (b) represents the normalized real and imaginary parts of the
conductivity, $\sigma_{\omega, q}'/\sigma_0$ and $\sigma_{\omega, q}''/\sigma_0$,
$\sigma_0 = 2 e^2 n_0/m^* V_T |q|$,
as functions of the normalized angular frequency, $\omega/qV_T$, for
$|\gamma_q| = 0,\,0.5,\,1$ and $1.5$ at $V_{dr}/V_T = 0.5$.
The right parts of these figures correspond to the case of $q>0,\,\omega>0$,
the left parts are for $q <0,\,\omega>0$. The apparent non-reciprocal frequency dispersion
of $\sigma_{\omega,q}$ is due to the electrons drift subjected to the
stationary field, $E_0$. In Figs.~\ref{fig1} (a) and (b), curves labeled by $1$ show normalized $\sigma_{\omega,q}'$
and $\sigma_{\omega,q}''$ calculated with the use of Eqs.~(\ref{sigma3}) and (\ref{sigma4}),
curves labeled by $2,\,3,\,4$ are calculated according to the result of
Eq.~(\ref{sigma2}). These curves demonstrate the importance of
the effect of the electric field, $E_0$, on the high-frequency electron dynamics.
Indeed, at a given $q$ an increase in $\gamma_q$ corresponds to a proportional
increase of $E_0$, this last leading to an oscillatory behavior of
both $\sigma_{\omega,q}'$ and $\sigma_{\omega,q}''$.
Moreover, since the real and imaginary parts of
$\sigma_{\omega,q}$ are of the same order of magnitude, a large phase shift, $\phi_{\omega}$, between the electric field $\tilde E$ and the current $\tilde j$ may appear; $\phi_{\omega}$  is strongly dependent on the frequency and can change a sign.

Fig.~\ref{fig1} (a) for $\sigma_{\omega,q}'$ shows that an increase of the field, $E_0$,
produces a widening of the Cerenkov-like region and {\it additional} high-frequency
regions with $\sigma_{\omega, q}' <0$. The oscillatory character of $\sigma_{\omega, q}'$
and the above mentioned frequency regions are better evidenced in Fig.~\ref{fig2} where
the density plot of $\sigma_{\omega, q}'$ is presented as a function of the  variables
$\{\omega/q V_T,\,\gamma_q\}$, which are dimensionless frequency and
dimensionless stationary field for a given $q$. The white regions correspond to $\sigma_{\omega, q}'>0$.
At $\gamma_q < 1$, i.e. at weak electric fields, we notice the Cerenkov-like region for small $\omega$
and an {\it additional} extensive high-frequency region with $\sigma_{\omega, q}' <0$.
At $\gamma_{q} > 1$, i.e. at large electric fields, there is an alternation of
regions with positive and negative $\sigma_{\omega, q}'$.

\begin{figure}[h]
  \includegraphics[width=0.45\textwidth]{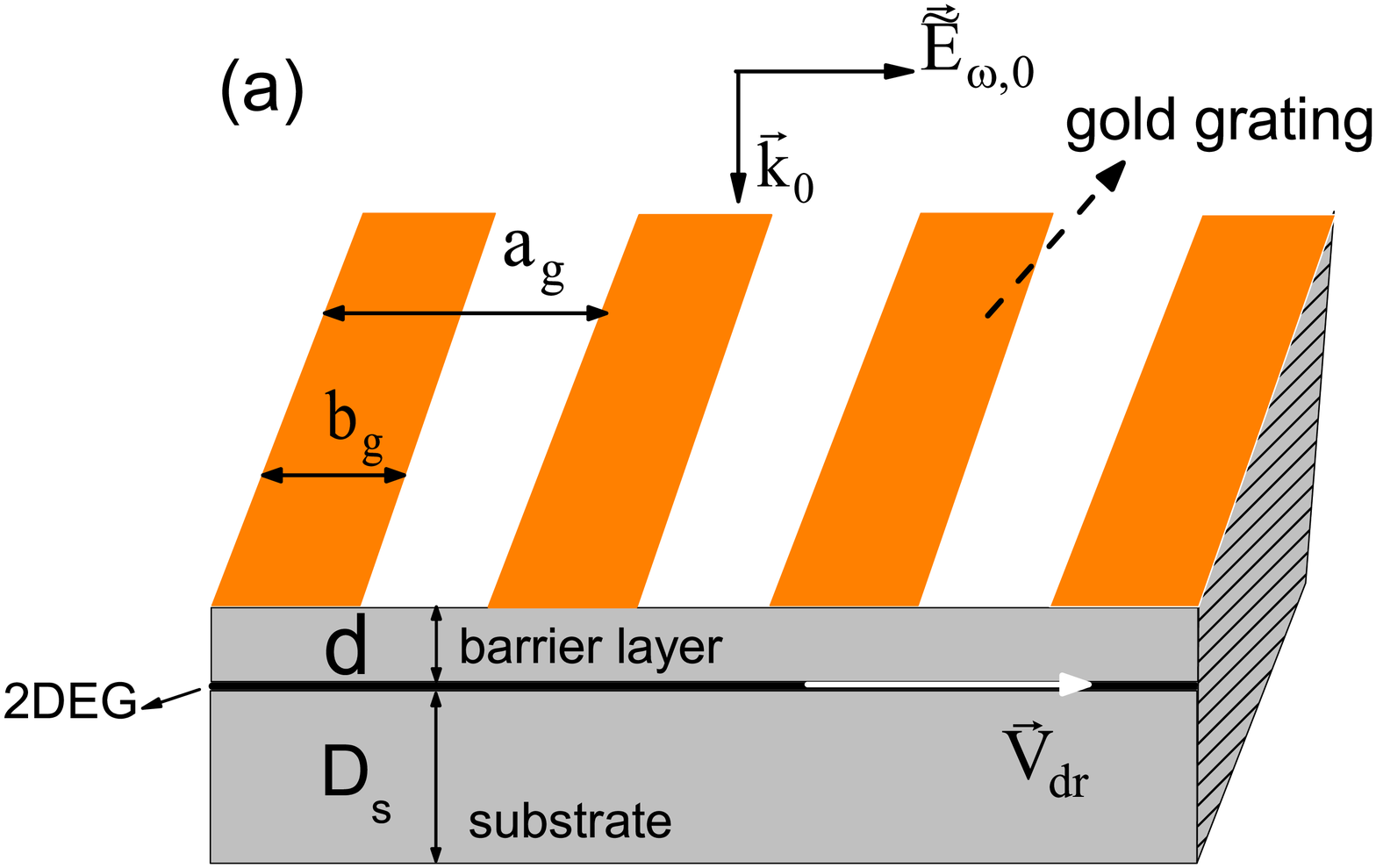}
  \includegraphics[width=0.45\textwidth]{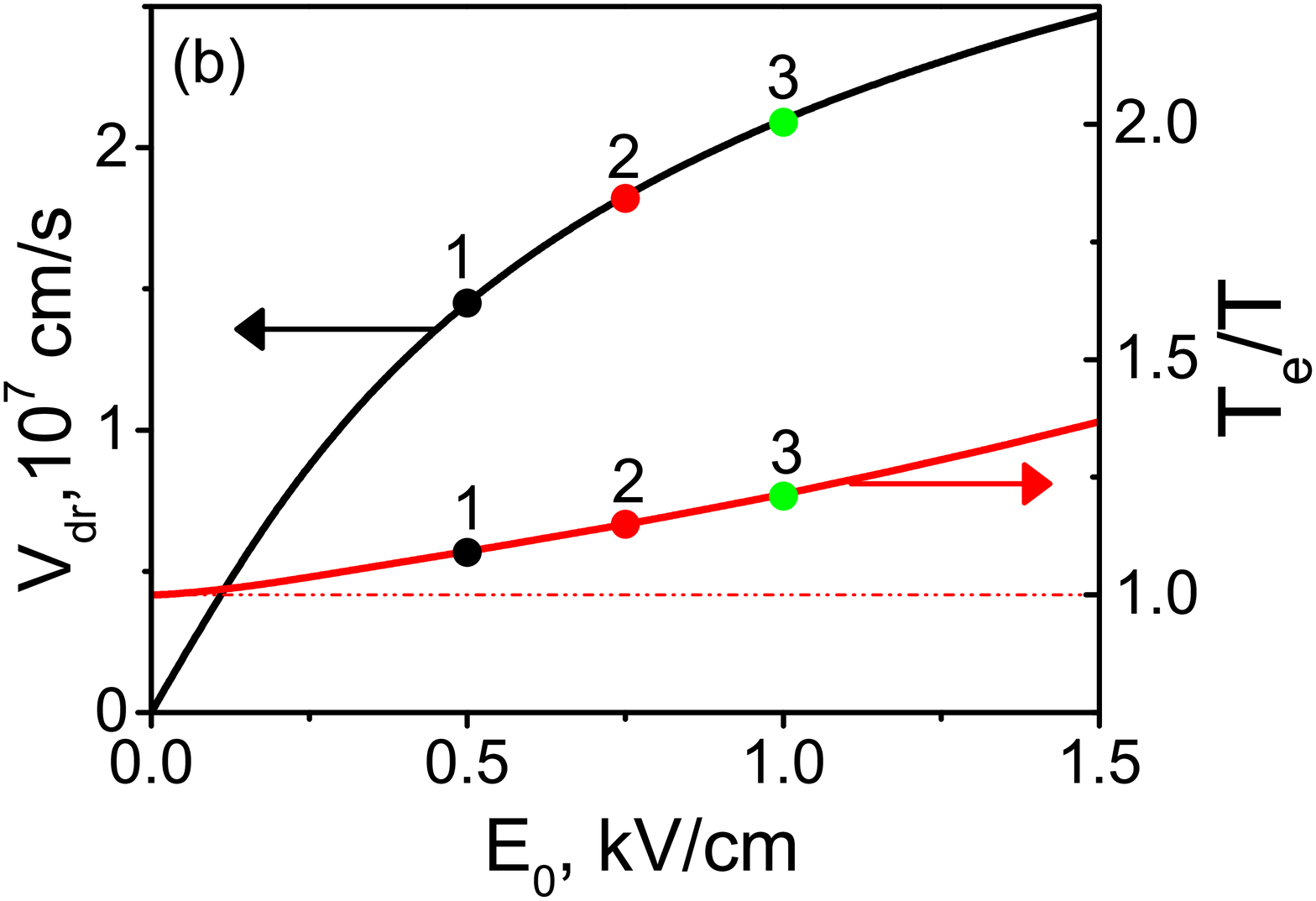}
   \caption{ (a): Schematic of the plasmonic structure with 2DEG.
(b): Dependencies of the electron temperature and drift velocity on the
applied electric field taken from Ref.~[\onlinecite{Kor2}]. The dots correspond to the electric fields for which the absorptivity is shown in Fig.~\ref{fig4}.}
  \label{fig3}
\end{figure}

The importance of a correct calculation of the high-frequency conductivity with
spatial dispersion can be illustrated with a practical example as follows.
Consider an AlGaAs/GaAs/AlGaAs
quantum well heterostructure covered by a submicron metallic grating.
A schematic of such a hybrid plasmonic structure is shown in Fig.~\ref{fig3}(a).
It is assumed that the structure is doped and there is a 2DEG in the GaAs quantum well layer.
A subwavelength metallic grating is placed in the vicinity of
the quantum well to provide a strong coupling of electron oscillations and
radiation under THz illumination of the plasmonic structure.
For the structure we set the following geometrical parameters: $a_{g}=200$ nm,
$b_{g}=160$ nm, $d=20$ nm, and  $D_{s}=2$ $\mu$m [see Fig.~\ref{fig3}(a)].
The sheet electron concentration is assumed to be $n_{0}=3\times10^{11}$~cm$^{-2}$.
A stationary lateral electric field,  $E_0$, applied to the quantum well layer
 induces a drift of the electrons. Calculations of $T_{e}$ and $V_{dr}$ in the GaAs quantum well
can be found elsewhere~\cite{Kor1, Kor2}. The dependencies $T_{e}(E_{0})$
and $V_{dr} (E_0)$ are presented in Fig.~\ref{fig3}(b) for a temperature of
$77$ K.

The $V_{dr} (E_0)$ function can be used to calculate the differential mobility, $\mu_d (E_0)\equiv d V_{dr}/d E_0$, and
the scattering time, $\tau_{sc} =m^* \mu_{d}(E_0)/e$, of hot electrons. The latter parameter allows us to estimate the criteria necessary for the collisionless approach: $\omega\tau_{sc}\gg 1,q l_{sc}\!=\!q \sqrt{2 k_B T_e/m^*}\tau_{sc}\!\!\gg\!\! 1$. For example, at $E_0=0.5$ kV/cm, we found  that  $\tau_{sc} = 0.65$ ps, $l_{sc} =1.3\times\,10^{-5}$ cm, i.e., for $3...4$ THz frequency range  we find that $\omega\,\tau_{sc} \approx 12...16$ and $q l_{sc}\approx 4$ at $q \approx \pi \times 10^{5}$ cm$^{-1}$ (characteristic wavevector corresponding to the grating period, $a_{g}$).

Based on the solution of the Maxwell's equation\cite{Popov, Kor4} and using Eq.~(\ref{sigma2}) together with the above mentioned data on $V_{dr}$ and $T_{e}$, we have calculated the spectral characteristics of the plasmonic structure, including transmittivity,  reflectivity and absorptivity. Absorptivity was calculated in usual way as 1 minus a sum of transmittivity and reflectivity\cite{THz-Ampl-2}.

\begin{figure}[h]
   \includegraphics[width=0.45\textwidth]{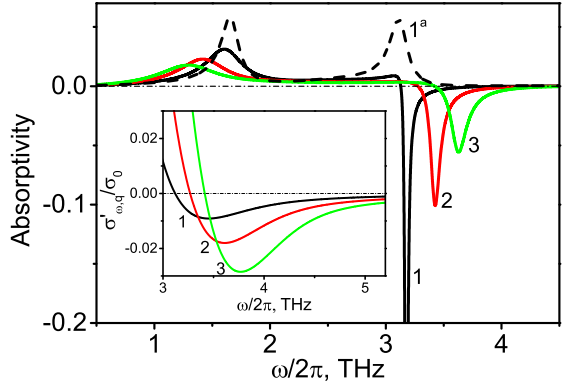}
  \caption{ Absorptivity spectra of the plasmonic structure at different applied electric fields indicated in the text.
Dashed curve is the absorptivity calculated at $T_e$ and $V_{dr}$ corresponding to $E_0=0.5$~kV/cm
and $\gamma_q=0$. Inset: normalized $\sigma_{\omega,q}'$ in the frequency range of $3...5$~THz, at $q =\pi \times 10^{5}$~cm$^{-1}$
and the same applied electric fields. }
  \label{fig4}
\end{figure}

In particular, the spectral
dependences of the absorptivity of THz waves calculated for three applied fields
$E_0 =0.5,\,0.75,\, 1 $~kV/cm (parameters $\gamma_q=0.21, 0.31, 0.39$, respectively) are shown in Fig.~\ref{fig4} by curves 1, 2, 3, respectively. The corresponding values of $T_e$ and $V_{dr}$ for these fields are indicated as dots in Fig.~\ref{fig3}(b).
For comparison, we present also the absorptivity calculated at $T_e$ and $V_{dr}$
corresponding to $E_0 = 0.5$ kV/cm, but setting $\gamma_q=0$, i.e., neglecting the electric field effect on the electron high-frequency dynamics (dashed line in Fig.~\ref{fig4}). The latter calculation put in evidence two peaks of absorption of THz waves related to the excitation of two plasmon waves propagating along the electron drift (higher frequency) and in the opposite
direction (lower frequency). In fact, more rigorous calculations  for $\gamma_q \neq 0$ show that
the lower frequency peak of absorption exists, though modified, but at higher frequencies (at $3...4$ THz) the absorptivity
becomes negative thus enhancing the intensity of THz radiation at the expense of the stationary field and current. In the corresponding frequency range, the sum of the amplitudes of the refracted and transmitted waves exceeds the incident wave amplitude.
For example, at $E_0=0.5$~kV/cm and resonant angular frequency $\omega/2\pi=3.18$~THz, the transmittivity and reflectivity
are equal to 1.01 and 0.52, respectively.  At higher fields, $E_{0}=0.75, \, 1$ kV/cm,
and corresponding resonant frequencies $\omega/2\pi=3.43,\, 3.63$ THz the transmittivity takes the values $0.65$ and $0.59$, respectively, with corresponding reflectivity values of $0.45$ and $0.47$. The  negative absorptivity corresponds to {\it additional} high-frequency
regions with $\sigma_{\omega, q}' <0$ (see inset in Fig.~\ref{fig4}). The emergence of negative absorptivity  corresponds to an energy transfer from the stationary field to the electromagnetic wave interacting with unstable plasmons modes.

\begin{figure}[h]
\centering
    \includegraphics[width=0.45\textwidth]{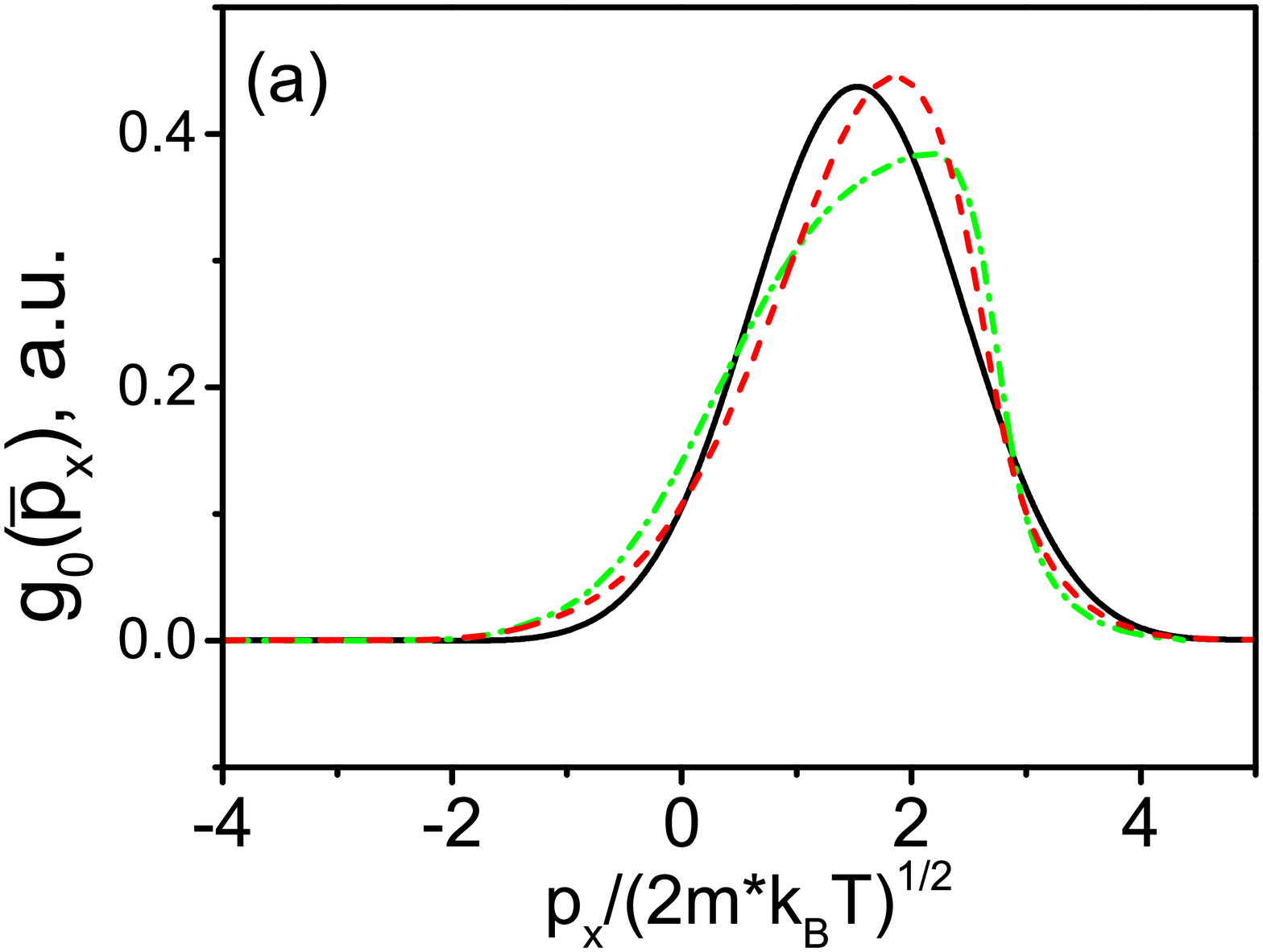}
    \includegraphics[width=0.45\textwidth]{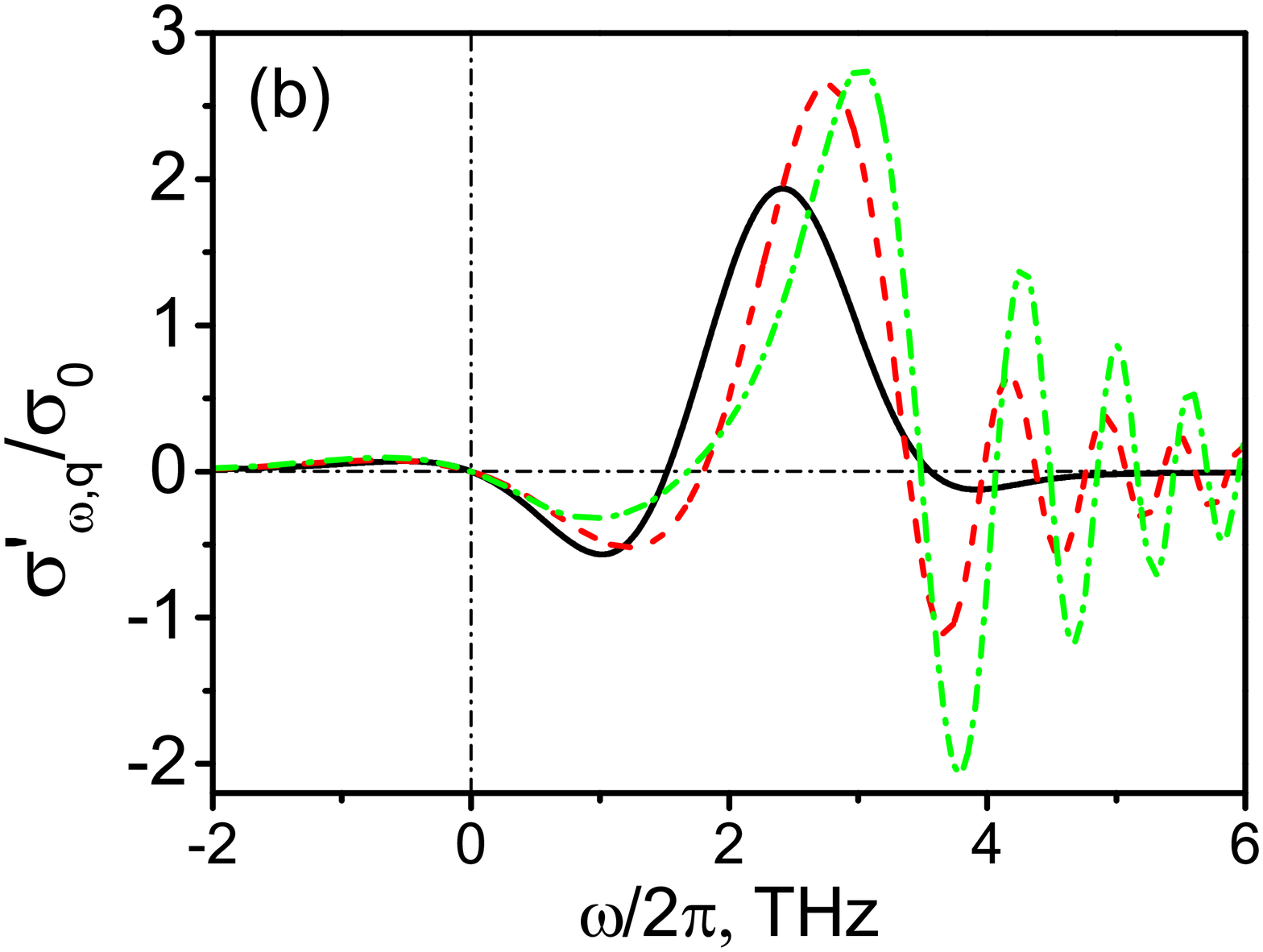}
  \caption{(a): The shifted   Maxwellian distribution function  $\bar{g}_{0}(p_{x})$
  (solid line), distribution functions obtained~\cite{Mosko} by the
  Monte-Carlo methods with and without e-e scattering (dashed and dashed-dot lines,
  respectively).
  (b): function $\sigma'_{\omega, q}$ calculated according Eq.(\ref{sigma_full})
  for the presented distribution functions. Wavevector $q_{x}$ is the same as in
  Fig.~\ref{fig4}.}
  \label{fig5}
\end{figure}

The above analysis was conducted for $\sigma_{\omega,q}$ with the use of the shifted
Maxwellian distribution, which facilitates the analytical
study of $\sigma_{\omega,q}$. Below, we demonstrate how the main features of $\sigma_{\omega,q}$
are reproduced for functions, ${g}_{0}(p_{x})$, obtained by the Monte-Carlo method.
We use the results of paper\cite{Mosko}, which were obtained by Monte-Carlo
simulations of electron transport in GaAs quantum wells at the parameters similar
to already used above. In particular, these results were reported for
the ambient temperature $T=60\,K$,  $E_0 = 1.2\,kV/cm$
and $n_{0}=2\times 10^{11}$ cm$^{-2}$ (i.e.,
{\it e-e} scattering does not completely control electron kinetics).
In Fig.~\ref{fig5}(a)) we show the shifted Maxwellian distribution (solid line)
with parameters $T_e=96\,K$ and $V_{dr} = 2.48\times 10^{7}\,cm/s$ (parameter $\gamma_q=0.47$), and
two functions $\bar{g}_{0}(p_{x})$ found~\cite{Mosko} with and without {\it e-e}
scattering (dashed and dashed-doted lines, respectively).
Though all three functions are seemingly quite similar and give the same mean energy
and drift velocity, the two latter functions  have
more sharp decrease at large momenta $p_x$.
Calculations of $\sigma'_{\omega,q}$ for these three functions $\bar{g}_{0}(p_{x})$
are presented in Fig.~\ref{fig5} (b). One can see that all discussed above
features of $\sigma'_{\omega,q}$ are well reproduced. Noticeable enhancement
of the oscillation behavior of $\sigma_{\omega,q}$ found for the functions obtained by
the Monte-Carlo method are due to more sharp decrease of the electron
stationary distribution at large $p_x$.

Summarizing, we have analyzed the high-frequency conductivity including
spatial dispersion for two-dimensional electrons subjected to a high
stationary electric field. We have taken into consideration the effects of the
stationary electric field on both the stationary electron distribution and the
high-frequency dynamics of the electrons. In the collisionless approximation we have found that
the high-frequency conductivity with spatial dispersion exhibits
the following features contrasting to the case of dissipative transport:
strong non-reciprocal effect, oscillatory behavior and a set of
frequency region with negative values of the real part of the
conductivity. If the 2DEG plasmon frequency is in one of these
regions, the current-driven 2DEG is unstable, i.e. the plasma oscillations will
grow in time and along the electron drift (so-called convective instability).
Under these conditions, an incident THz wave can be amplified in
a properly designed hybrid plasmonic structure. Similarly, in a hybrid system composed of electrostatically coupled quantum
well and a polarizable nanoparticle (a quantum dot, a molecule, etc.),
the electron drift in the stationary electric field will provide
 excitation and instability of this hybrid system if the dipole oscillation frequency
of the nanoparticle is in one of the discussed frequency
regions\cite{Kukhtaruk1}.

We suggest that the discovered features associated with the electron response
to high-frequency and spatially nonuniform electromagnetic fields are
of general character. The obtained results  can be used for
the refining of  near-field THz microscope techniques and the development of THz devices with a lateral nanostructuring.

{\it This work is supported by the German Federal Ministry of
Education and Research (BMBF Project 01DK17028).}

\end{document}